\RequirePackage{fix-cm}
\documentclass[twocolumn,epjc3,final]{svjour3}  
\smartqed  
\RequirePackage{graphicx}

\usepackage{xcolor}
\usepackage{comment}
\usepackage{amsmath, amsfonts,amssymb}
\usepackage{xparse}
\usepackage{mathrsfs}
\usepackage{breqn}
\usepackage{braket}

\newcommand*{\diag}{\operatorname{diag}}

\usepackage[numbers,sort&compress]{natbib}
\usepackage[colorlinks,citecolor=blue,urlcolor=blue,linkcolor=blue]{hyperref}
\journalname{Eur. Phys. J. C}
\begin{document}

\title{Decomposition of the connection in affine models of gravity\thanksref{t1}}
\subtitle{Can the connection tell us something about the metric?}

\titlerunning{Decomposition of the affine connection}

\author{Oscar Castillo-Felisola\thanksref{e1}
  \and
  Bastian Grez\thanksref{addr1,addr3}
  \and
  Aureliano Skirzewski\thanksref{e2}
  \and
  Jefferson Vaca-Santana\thanksref{addr3,addr1}
}

\thankstext{t1}{This work has been fund by ANID PIA/APOYO AFB230003 (Chile) and FONDECYT Grant 1230110 (Chile)}
\thankstext{e1}{e-mail: o.castillo.felisola:at:proton.me}
\thankstext{e2}{e-mail: askirz:at:gmail.com}


\institute{Departamento de F\'isica, Universidad T\'{e}cnica Federico Santa Mar\'ia, Casilla 110-V, Valpara\'iso, Chile \label{addr1}
  \and
  Centro Cient\'ifico Tecnol\'ogico de Valpara\'iso, Casilla 110-V, Valpara\'iso, Chile \label{addr2}
  \and
  Instituto de F\'isica, Pontificia Universidad Cat\'olica de Valpara\'iso, Av. Brasil 2950, Valpara\'iso, Chile \label{addr3}
  \and
  Instituto de F\'isica, Facultad de Ciencias, Igu\'a 4225, esq. Mataojo, 11400 Montevideo, Uruguay \label{addr5}
}

\date{Received: date / Accepted: date}

\maketitle

\begin{abstract}
  In physics geometrical connections are the mean to create models with local symmetries (gauge connections), as well as general diffeomorphisms invariance (affine connections). Here we study the irreducible tensor decomposition of connections on the tangent bundle of an affine manifold as used in the polynomial affine model of gravity \cite{castillo-felisola24_polyn_affin_model_gravit}. This connection is the most general linear connection, which allows us to build metric independent, diffeomorphism invariant models. This set up includes parts of the connection that are associated with conformal and projective transformations.
\end{abstract}

\section{Introduction}

The absolute calculus was developed during a period ranging from the last decades of the XIX century until the first couple of decades of the XX century, by a handful of people like Christoffel, Levi-Civita, Ricci-Curbastro and others \cite{christoffel69_ueber_trans_homog_differ,ricci00_method_de_calcul_absol,levi-civita16_nozion_di_paral_una}, attempting to extend the notion of calculus to non-Euclidean geometries, which were being classified by Klein's Erlangen Program \cite{klein08_compar_review_recen_resear_geomet,scholz22_h} (the completion of the program was due to E. Cartan, see Ref.~\cite{rosenfeld88_euclid} for a historical perspective).

At the time, mathematicians understood that two structures could be used to define the absolute calculus, Riemann's groundform (also known as the metric tensor field) and the affine connection \cite{eisenhart27_non_rieman,weyl22_space,schouten13_ricci}. These structures play the role of the compass and ruler in Euclidean geometry, allowing us to introduce the notions of distance (and angles) and parallelism, respectively.

The point of view of Minkowski of Einstein's restricted theory of relativity \cite{einstein20_princ_relat} extended Riemannian geometry to semi-Riemannian (or Lorentzian) ones, providing the ground structure for General Relativity \cite{einstein15_grund_allgem_relat_und_anwen,einstein15_zur_allgem_relat,einstein16_grund_allgem_relat}. However, it was noticed by Einstein himself \cite{einstein23_theor_affin_field,einstein23_zur_affin_feldt} and Eddington \cite{eddington23} that the gravitational interaction can be modeled using the affine connection, formulating the first purely affine model of gravity. Further affine models of gravity have been proposed by Kijowski~\cite{kijowski78_new_variat_princ_gener_relat,kijowski79,ferraris81_gener_relat_is_gauge_type_theor,ferraris82_equiv_relat_theor_gravit,kijowski07_univer_affin_formul_gener_relat}, Pop{\l}awski~\cite{poplawski07_nonsy_purel_affin,poplawski07_unified_purel_affin_theor_gravit_elect,poplawski09_gravit_elect,poplawski14_affin_theor_gravit}, Azri~\cite{azri17_affin_inflat,azri18_cosmol_implic_affin_gravit,azri18_induc_affin_inflat,azri19_induc_gravit_from_connec_scalar_field} and Castillo-Felisola~\cite{castillo-felisola15_polyn_model_purel_affin_gravit,castillo-felisola18_einst_gravit_from_polyn_affin_model,castillo-felisola18_beyond_einstein,castillo-felisola20_emerg_metric_geodes_analy_cosmol,castillo-felisola24_polyn_affin_model_gravit} and their collaborators.

While semi-Riemannian geometries inherit a \emph{natural} (and unique) affine connection, defined in terms of the metric and its derivatives---the Levi-Civita connection, which is symmetric and compatible with the metric structure \cite{choquet-bruhat89_analy,nakahara05_geomet_topol_physic}, in metric-affine geometries these structures are independent \cite{eisenhart27_non_rieman,weyl22_space,schouten13_ricci}. However, in the later scenario one can decompose the affine connection into three parts: (i) the Levi-Civita connection, which depends solely on the metric; (ii) the contorsion, which is built from the non-symmetric part of the connection called torsion; and, (iii) the deflection, associated to the failure of satisfying the metricity condition.

In affine geometries, the above decomposition of the connection is simpler, it can decomposed into symmetric and non-symmetric components, but the lack of metric forbids further decomposition (with the exception of traces). Specifically, the absence of a metric makes it impossible to separate the Levi-Civita and nonmetricity components of the connection, nor build the contorsion and deflection tensor. However, one could take any nondegenerate, symmetric \(\binom{0}{2}\)-tensor field as an auxiliary ``metric'' and use it to decompose the affine connection as in metric-affine geometries \cite{schouten13_ricci}.

In all of the aforementioned geometries (metric, affine and metric affine), the invariance under diffeomorphisms might be achieved when building gravitational models. However, it is more difficult to test the background independence of the model, in particular when the intuition about distances introduced by the metric is off the table.

The purpose of this article is to analyze the behavior of the various components of the affine connection, particularly when it is compatible with certain symmetry group, under gauge transformations (deformations) of the auxiliary ``metric''.

The article is organized as follows. In Sec.~\ref{sec:work-decomp-conn} we give a brief overview of the generic decomposition of the affine connection, and analyze the response of its components under variations in the choice of metric. Our purpose then is to focus our attention on connections compatible with certain symmetry groups, specifically \(SO(n-1,\mathbb{R})\) and \(ISO(n-1,\mathbb{R})\), which are the groups behind spherically symmetric and cosmological \(n\)-dimensional configurations, respectively. However, we note that the decomposition analyses differ depending on the dimension of the set-up. In Sec.~\ref{sec:connections-four}, we work out the detailed decomposition of the connection in dimension four and higher, while the two-dimensional and three-dimensional cases are developed in Secs.~\ref{sec:conn-two-dimens} and~\ref{sec:conn-three-dimens}. Some conclusions are drawn in Sec.~\ref{sec:concluding-remarks}.

\section{Working out the decomposition of the connection\label{sec:work-decomp-conn}}

In general, no metric is necessary to introduce a diffeomorphism invariant model to describe purely affine models of gravity. However, the use of a metric is certainly the most intuitive way to describe what we see in a physical scenario.

Within the general framework of the polynomial affine model of gravity, the connection can be described using an auxiliary metric (as proposed in Ref.~\cite{schouten13_ricci}) but in general there is a gauge symmetry that connects one choice of metric with another, through inhomogeneous transformations of the nonmetricity. A detailed explanation is probably too evasive, so, let us assume that we can decompose the role of a symmetric connection $\Gamma_{\mu}{}^{\lambda}{}_{\nu}$ into its Riemannian part,
\begin{equation}
  \Gamma_{\mu}{}^{\lambda}{}_{\nu}(g)=\frac{1}{2} g^{\lambda\kappa}(\partial_{\mu} g_{\nu\kappa} + \partial_{\nu} g_{\mu\kappa} - \partial_{\kappa} g_{\mu\nu})
\end{equation} 
and its non-Riemannian parts, which relate to the nonmetricity through\footnote{The decomposition can be obtained using the Young decomposition of tensors, \([1] \otimes [1] \otimes [1] = [3] \oplus [2,1] \oplus [2,1] \oplus [1,1,1]\). The tensor \(\hat{S}\) belongs to the subspace \([3]\), while \(\hat{Y}\) belongs to the symmetric part of the two spaces \([2,1]\).}
\begin{equation}
  \nabla^{\Gamma}_\lambda g_{\mu\nu}
  =
  \hat Y_{\lambda\mu\nu}
  + \hat S_{\lambda\mu\nu},
\end{equation} 
where 
\begin{equation}
  \hat Y_{\lambda\mu\nu} = \frac{1}{2} (\hat Y_{[\lambda\mu]\nu} + \hat Y_{[\lambda\nu]\mu})
\end{equation}
and
\begin{equation}
\hat S_{\lambda\mu\nu} = \hat S_{(\lambda\mu\nu)}. 
\end{equation}

In the presence of torsion, the full connection $\hat\Gamma_{\mu}{}^{\lambda}{}_{\nu}$ can be reexpressed in terms of the torsion tensor and the symmetric connection 
\begin{dmath}
  \hat\Gamma_{\mu}{}^{\lambda}{}_{\nu}=\Gamma_{\mu}{}^{\lambda}{}_{\nu}+\frac{1}{2} T^{\lambda}{}_{\mu\nu} .
\end{dmath}
The torsion itself can be decomposed as
\begin{dmath}
  \frac{1}{2}T^{\lambda}{}_{\mu\nu}=\mathcal{B}_{\mu}{}^{\lambda}{}_{\nu}+\mathcal{A}_{[\mu}\delta^\lambda_{\nu]}
\end{dmath}
where $\mathcal{B}_{\mu}{}^{\lambda}{}_{\nu}$ is traceless. If the connection is metric compatible, it is reexpressed in terms of the contorsion tensor
\begin{equation}
  \hat\Gamma_{\mu}{}^{\lambda}{}_{\nu}=\Gamma_{\mu}{}^{\lambda}{}_{\nu}(g)+K^{\lambda}{}_{\mu\nu}
\end{equation}
where
\begin{equation}
  K^{\lambda}{}_{\mu\nu} = \frac{1}{2} \left( T^{\lambda}{}_{\mu\nu} + T_{\mu\nu}{}^{\lambda} + T_{\nu\mu}{}^{\lambda} \right),
\end{equation}
but in this context, the separation of the symmetric part of the connection and the contorsion cannot be justified.

Invariance under the choice of metric implies that the connection $\hat\Gamma_{\mu}{}^{\lambda}{}_{\nu}$ is invariant under infinitesimal transformations 
\begin{equation}
  \label{metricGeneralTransformations}
  g_{\mu\nu} \to g'_{\mu\nu} = g_{\mu\nu} + s_{\mu\nu},
\end{equation} 
where $s_{\mu\nu}$ is symmetric. The torsion tensor is invariant. In order for the connection to be invariant,
\begin{equation}
  \Gamma_{\mu}{}^{\lambda}{}_{\nu}\longrightarrow \Gamma'_{\mu}{}^{\lambda}{}_{\nu} = \Gamma_{\mu}{}^{\lambda}{}_{\nu},
\end{equation} 
the components of the nonmetricity have to transform as 
\begin{equation}
  \hat Y_{\lambda\mu\nu}
  \to
  \hat Y'_{\lambda\mu\nu}
  =
  \hat Y_{\lambda\mu\nu}
  + \frac{2}{3} \left(\nabla^\Gamma_{[\lambda} s_{\mu]\nu}+\nabla^\Gamma_{[\lambda}s_{\nu]\mu}\right)
\end{equation}
and
\begin{equation}
  \hat S_{\lambda\mu\nu}
  \to
  \hat S'_{\lambda\mu\nu}
  =
  \hat S_{\lambda\mu\nu}
  - \frac{1}{2}\nabla^\Gamma_{(\lambda}s_{\mu\nu)}.
\end{equation}

In spite of the general invariance under metric change observed in the polynomial affine model of gravity, it turns convenient to describe its parts in the more conventional language provided by the choice of a metric, mainly in order to compare what we observe in the solutions of the polynomial affine model of gravity with Einstein's gravity.
Thus, we will propose a given metric such that the symmetric part of the affine connection is 
\begin{equation}
  \Gamma_{\mu}{}^{\lambda}{}_{\nu}
  =
  \Gamma_{\mu}{}^{\lambda}{}_{\nu}(g)
  + \hat S^\lambda{}_{\mu\nu}
  + \hat Y^\lambda{}_{\mu\nu}.
\end{equation} 
Additionally, we can decompose $\hat S^\lambda{}_{\mu\nu}$ and $\hat Y^\lambda{}_{\mu\nu}$ further by writing the trace separately 
\begin{equation}
  \label{ConnectionDecomposition}
  \Gamma_{\mu}{}^{\lambda}{}_{\nu} - \Gamma_{\mu}{}^{\lambda}_{\nu}(g) = S{}^\lambda{}_{\mu\nu}+Y{}^\lambda{}_{\mu\nu}+V^\lambda g_{\mu\nu}+2W_{(\mu}\delta^\lambda_{\nu)},
\end{equation} 
where $S^\lambda{}_{\mu\nu}$ and $Y^\lambda{}_{\mu\nu}$ are, assumed to be traceless, and in \(n\) dimensions $S^\lambda{}_{\mu\nu}$ has
\begin{equation}
  [S_{\lambda\mu\nu}] = \frac{n(n+1)(n+2)}{3!} - n
\end{equation}
independent components, while $Y^\lambda{}_{\mu\nu}$ has
\begin{equation}
  [Y_{\lambda\mu\nu}] = \frac{n(n+1)(n-1)}{3} - n.
\end{equation}

As a side note, notice that the expression above is not polynomial in the metric, but we use it as a gauge choice for the sake of gaining some understanding of the space of solutions.

\section{Connections in four or more dimensions\label{sec:connections-four}}

In this section, we shall work out the decomposition of the irreducible components of the symmetric affine connection, considering that the background space has dimension four or higher. 

\subsection{Transverse traceless symmetric three-tensor}

After decomposing the most general symmetric connection in terms of the Levi-Civita symbols and nonmetricity, it becomes clear that the number of components in the parts exceeds by $n(n+1)/2$ the number of components in an affine connection:
\begin{dmath}
  [g_{\mu\nu}]
  + [S_{\mu\nu\lambda}]
  + [Y_{\mu\nu\lambda}]
  + [W_\mu]
  + [V^\mu]
  - [\Gamma_\mu{}^\lambda{}_\nu]
  =
  \frac{n(n+1)}{2},
\end{dmath}
that is assuming that the degrees of freedom in the Christoffel symbols are given by the metric
$$[\Gamma_\mu{}^\lambda{}_\nu(g)]=[g_{\mu\nu}].$$

Therefore, we could assume that either the metric should not count as a degree of freedom or that we could fix the gauge that allows us to change the metric to reduce the degrees of freedom of other fields.

In fact, defining $S^{TT}_{\mu\nu\lambda}$ as the part of $S_{\mu\nu\lambda}$ that satisfies $\nabla^\mu S^{TT}_{\mu\nu\lambda} = 0$, we realize its components count is
\begin{dmath}
  [S^ {TT}_{\mu\nu\lambda}]
  =
  \frac{(n-1)(n-2)(n+3)}{6}
  =
  [S_{\mu\nu\lambda}]
  - ([g_{\mu\nu}] - 1).
\end{dmath}
This last identity suggests that we can extract the metric degrees of freedom from $S_{\mu\nu\lambda}$ and include its transverse part. As for the extra degree of freedom that remains, conformal transformations can be used later to understand the kinematics of the model and further fix either the metric, $V^\mu$ or $W_{\mu}$.

\paragraph{Cosmological decomposition.---}
An $S_{\mu\nu\lambda}$ that is compatible with the cosmological principle can be written in terms of two symmetric objects: a timelike vector, $T_\mu=(1,\vec 0)$, and the metric of the spatial \(n-1\) submanifold, $s_{ij}$, both invariant under rotations and translations. We define the cosmological metric as 
\begin{equation}
  g_{\mu\nu} dx^\mu \otimes dx^\nu= - N^2 \, dt^2 + a^2 \, s_{ij} dx^i \otimes dx^{j},
\end{equation}
and the \(S\) tensor can be expressed as
\begin{equation}
  S_{\mu\nu\lambda} = A \left(T_\mu T_\nu T_\lambda + \frac{3}{n-1} \frac{a^2}{N^2} \delta_{(\mu}^i \delta_\nu^j T_{\nu)} s_{ij}\right),
\end{equation}
or explicitly in components,
\begin{dmath}
  S_{ttt}=A
\end{dmath}
and
\begin{dmath}
  S_{tij} = \frac{1}{n-1} \frac{a^2}{N^2} s_{ij} A.
\end{dmath}
The tensor $S_{\mu\nu\lambda}$ is said to be transverse if $\nabla^\mu S_{\mu\nu\lambda}=0$, thus 
\begin{dmath}
  \nabla^\mu S_{\mu\nu\lambda}=\frac{1}{\sqrt{-g}}\partial_\rho(\sqrt{-g}g^{\rho\mu}S_{\mu\nu\lambda})-g^{\rho\mu}\Gamma_{\rho}{}^\sigma{}_\nu(g) S_{\mu\sigma\lambda}-g^{\rho\mu}\Gamma_{\rho}{}^\sigma{}_\lambda(g) S_{\mu\nu\sigma}
\end{dmath}
using the components of the Levi-Civita connection,
\begin{align}
  \Gamma _{k}{}^{j}{}_t(g) = \frac{\dot a}{a} \delta _{k}^{j}, \\
  \Gamma_i{}^t{}_j(g)      = \frac{a\dot a}{N^2} s_{ij}, \\
  \Gamma _{t}{}^{t}{}_t(g) = \frac{\dot N}{N},
\end{align}
it follows that
\begin{dmath}
  \nabla ^{\mu}S_{\mu ij}
  = - \frac{1}{Na^{3}} \partial_{t}\left(\frac{a^{3}}{N} S_{tij}\right)
  = - \frac{s_{ij}}{(n-1) N a^{3}} \partial_{t} \left(\frac{a^5}{N^3} A\right),
\end{dmath}
\begin{dmath}
  \nabla ^{\mu}S_{\mu i t}
  =\frac{1}{a^2\sqrt{s}}\partial_i(\sqrt{s})\frac{1}{n-1}\frac{a^2}{N^2}A
  -\frac{1}{a^2}\Gamma_{j}{}^j{}_i(g) \frac{1}{n-1}\frac{a^2}{N^2}A
  =0,
\end{dmath}
and
\begin{dmath}
  \nabla ^{\mu}S_{\mu tt}=\frac{1}{\sqrt{-g}}\partial_t(\sqrt{-g}g^{tt}S_{t tt})
  -2g^{\rho\mu}\Gamma_{\rho}{}^\sigma{}_t(g) S_{\mu\sigma t}
  =-\dfrac{1}{Na^{3}}\partial _{t}\left( \frac{a^3}{N}S_{ttt}\right) 
  +\dfrac{2\dot N}{N^3} S_{ttt}
  -\dfrac{2 \dot a}{a^3}s^{ij}S_{tij}
  =-\dfrac{1}{Na^{3}}\partial _{t}\left( \frac{a^3}{N}A\right) 
  +\dfrac{2\dot N}{N^3}A
  -\dfrac{2 \dot a}{a N^2}A
  =-\dfrac{N}{a^{5}}\partial _{t}\left( \frac{a^{5}}{N^3}A\right).
\end{dmath}
A nontrivial $A$ can be fixed by the geometry, \(A \propto \frac{N^3}{a^5}\), therefore it is not a degree of freedom by itself. Finally, we write the most general solution,
\begin{equation}
  S_{ttt} = \sigma\frac{N^3}{a^5} \text{, and }S_{tij}=\frac{1}{n-1}s_{ij}\sigma\frac{N}{a^3}.
\end{equation}

This non-zero result may sound strange, but when we compare it to the solutions of other gauge fixings in the literature, we find that this is usually called a residual gauge \cite{bertlmann96_anomal_quant_field_theor}, which just speaks of the inability of the gauge fixing condition to get rid of all the physically insignificant content in the field. Trusting this last assessment, we may set it to zero; however,  this field appears directly on the geodesics, and that makes it physically relevant. 

\paragraph{Static spherical decomposition.---}
In spherically symmetric spaces there is an $n-2$ dimensional sphere with metric $s_{ab}$, where the indices \(a,b = 2, \cdots, n\), any tensor with angular components has to be proportional to this metric or the \((n-2)\)-dimensional skew symmetric Levi-Civita $\sqrt{s}\epsilon^{a_1 \cdots a_{d-2}}$ (not useful to describe $S^{TT}_{\mu\nu\lambda}$). In the time-like direction, there is homogeneity and time reversal symmetry. Thus, we get an invariant time-like vector $T_\mu=(1,0,\vec 0)$, which has to be used in pairs, and a rotationally invariant radial vector $R_\mu=(0,1,\vec 0)$ with no parity. The most general ansatz is a traceless version of 
\begin{dmath}
  S^{TT}_{\mu\nu\lambda}
  =
  A \, T_{(\mu}T_\nu R_{\lambda)}
  + B \, R_{(\mu}R_\nu R_{\lambda)}
  + C \, R_{(\mu}\delta^a_\nu \delta^b_{\lambda)} s_{ab}.
\end{dmath}
Using the metric
\begin{equation}
  g_{\mu\nu} dx^\mu \otimes dx^\nu= - F \, dt^2 + \frac{G}{F} \, dr^2 + r^2 \, s_{ab} dx^a \otimes dx^{b},
\end{equation}
the tracelessness condition of  \(S^{TT}_{\mu\nu\lambda}\) implies that
\begin{equation}
  C = -\frac{r^2}{n-2} \left(A \, T^2 + 3 B \, R^2 \right),
\end{equation}
hence
\begin{dmath}
  S^{TT}_{\mu\nu\lambda}
  =
  A \, T_{(\mu}T_\nu R_{\lambda)}
  + B \, R_{(\mu}R_\nu R_{\lambda)}
  + \frac{r^2}{n-2} \left(F \, A  - \frac{3G}{F} B\right) R_{(\mu}\delta^a_\nu \delta^b_{\lambda)} s_{ab}.
\end{dmath}
Given that the nontrivial components of the Levi-Civita connection are
\begin{equation}
  \begin{aligned}
    \Gamma_{t}{}^{t}{}_{r}(g) & = \Gamma_{r}{}^{t}{}_{t}(g) = \frac{F'}{2F}, \\
    \Gamma_{t}{}^{r}{}_{t}(g) & = \frac{F'F}{2G}, \\
    \Gamma_{r}{}^{r}{}_{r}(g) & = \frac{F}{2G} \left(\frac{G}{F}\right)', \\
    \Gamma_{a}{}^{r}{}_{b}(g) & = - \frac{rF}{G} s_{ab} ,\\
    \Gamma_{r}{}^{a}{}_{b}(g) & = \Gamma_{b}{}^{a}{}_{r}(g) = \frac{1}{r} \delta^a_b, \\
    \Gamma_{a}{}^{c}{}_{b}(g) & = \Gamma_{a}{}^{c}{}_{b}(s),
  \end{aligned}
\end{equation}
we can evaluate the expression
\begin{dmath}
  \nabla^\mu S^{TT}_{\mu\nu\lambda}
  =
  \frac{1}{\sqrt{-g}}\partial_\rho(\sqrt{-g}g^{\rho\mu}S_{\mu\nu\lambda})-g^{\rho\mu}\Gamma_{\rho}{}^\sigma{}_\nu(g) S_{\mu\sigma\lambda}-g^{\rho\mu}\Gamma_{\rho}{}^\sigma{}_\lambda(g) S_{\mu\nu\sigma}
  =
  0.
\end{dmath}
Hence, the vanishing condition for 
\begin{dmath}
  \nabla^\mu S^{TT}_{\mu tt}=\frac{1}{\sqrt{G}r^{n-2}}\partial_r\left(\frac{r^{n-2}FA}{3\sqrt{G}}\right),
\end{dmath}
implies that $\frac{r^{n-2}FA}{\sqrt{G}}$ is constant, say \(A_0\). Similarly,
\begin{dmath}
  \nabla^\mu S^{TT}_{\mu t r} = 0,
\end{dmath}
\begin{dmath}
  \nabla^\mu S^{TT}_{\mu t a} = 0,
\end{dmath}
\begin{dmath}
  \nabla^\mu S^{TT}_{\mu r a}
  =
  \frac{1}{r^2\sqrt{s}} \partial_c(\sqrt{s}) s^{cb}S_{a b r}
  -g^{lj}\Gamma_{l}{}^k{}_i(g) S_{jk r}
  =0,
\end{dmath}
\begin{dmath}
  \nabla^\mu S^{TT}_{\mu a b}
  =
  \frac{1}{\sqrt{G}r^{n-2}} \partial_r \left(r^{n-2} \frac{F}{\sqrt{G}} S_{a b r}\right)
  - 2 g^{c d} \Gamma_{d}{}^r{}_{(b}(g) S_{a) c r}
  - 2 g^{rr} \Gamma_{r}{}^c{}_{(b}(g) S_{a) r c}
  =
  \frac{s_{ab}}{3 \sqrt{G} r^{n-2}} \partial_r
  \left(r^{n-2} \frac{F}{\sqrt{G}} C\right) ,
\end{dmath}
from which $r^{n-2} \frac{F}{\sqrt{G}} C = C_0$ is constant, and finally
\begin{dmath}
  \nabla^\mu S^{TT}_{\mu rr}
  =
  \frac{1}{\sqrt{-g}} \partial_r(\sqrt{-g}g^{rr}S_{rrr})
  - 2 g^{rr} \Gamma_{r}{}^r{}_r(g) S_{r r r}
  - 2 g^{tt} \Gamma_{t}{}^t{}_r(g) S_{t t r}
  - 2 g^{bc} \Gamma_{c}{}^a{}_r(g) S_{abr}
  =
  \frac{1}{r^{n-2}\sqrt{G}} \partial_r(r^{n-2} \sqrt{G} \frac{F}{G} B)
  - 2 \frac{F}{G} \frac{F}{2G} \left(\frac{G}{F}\right)' B
  - 2 \frac{-1}{F} \frac{F'}{2F} \frac{A}{3} 
  - \frac{2}{r^2} s^{bc} \frac{1}{r} \delta_c^a \frac{s_{ab}}{3} C
  =
  \frac{\sqrt{G}}{r^{n-2}F} \partial_r(r^{n-2}\frac{F^2}{\sqrt{G^3}}B)
  + \frac{F'}{F^2} \frac{\sqrt{G}A_0}{3r^{n-2}F} 
  - \frac{2(n-2)}{3} \frac{\sqrt{G}C_0}{r^{n+1}F}
  =
  \frac{\sqrt{G}}{r^{n-2}F} \left[\partial_r \left( r^{n-2} \frac{F^2}{\sqrt{G^3}} B \right) + \frac{F'A_0}{3F^2} - \frac{2(n-2)}{3} \frac{C_0}{r^{3}} \right].
\end{dmath}
Hence, $\frac{r^{n-2}F^2B}{\sqrt{G^3}} = B_0 + \frac{A_0}{3F} - \frac{n-2}{3r^2} C_0,$ which is compatible with the tracelessness condition if $B_0=0$. This, as in the cosmological case, indicates that there are no proper degrees of freedom associated to $S^{TT}_{\mu\nu\lambda}$, which doesn't mean these terms should be absent from the equations of motion or the geodesics, as there may be residual gauge determined exclusively by the metric. Still, a translation from the degrees of freedom of an arbitrary connection that has the required symmetries, to the degrees of freedom from the metric and remaining nonmetricity components is still feasible.

\subsection{Transverse traceless mixed tensor}

The tensor decomposition of the connection could be used to separate transverse from longitudinal degrees of freedom as
\begin{equation}
  [Y_{\mu\nu\lambda}] - [Y^{TT}_{\mu\nu\lambda}]
  =[\sigma_{\mu\nu}]+[\omega^T_{\mu\nu}],
\end{equation}
where $\sigma_{\mu\nu}$ is a symmetric tensor and $\omega^T_{\mu\nu}$ is a transverse antisymmetric tensor. This may be used to write 
\begin{dmath}
  Y_{\lambda\mu\nu}=Y^{TT}_{\lambda\mu\nu}+\left(P_{\lambda\mu\nu}^{\lambda'\mu'\nu'}-\widetilde{P}_{\lambda\mu\nu}^{\lambda'\mu'\nu'}\right)\left(\nabla_{\lambda'}\sigma_{\mu'\nu'}+\nabla_{\mu'}\omega^T_{\nu'\lambda'}\right),
\end{dmath}
where
\begin{equation}
  P_{\lambda\mu\nu}^{\lambda'\mu'\nu'} = \frac{4}{3}\delta_{\lambda}^{[\lambda'}\delta_{(\mu}^{\mu']}\delta_{\nu)}^{\nu'},
\end{equation}
is the projector into the symmetric representation of Young type \([2,1]\), while the trace is removed with the  projector
\begin{equation}
  \widetilde{P}_{\lambda\mu\nu}^{\lambda'\mu'\nu'}=\frac{4}{3(n-1)}\left(g_{\mu\nu}\delta_{\lambda}^{[\lambda'}g^{\mu']\nu'}-g_{\lambda(\mu}\delta_{\nu)}^{[\lambda'}g^{\mu']\nu'}\right),
\end{equation}
and $\sigma_{\mu\nu}$ may be further decomposed.  It may be possible to fix $\sigma_{\mu\nu}=e^{2\phi}g_{\mu\nu}$ by choosing a given gauge, but in general it would not be possible to set $S_{\lambda\mu\nu}=S^{TT}_{\lambda\mu\nu}$ and $\sigma_{\mu\nu}=e^{2\phi}g_{\mu\nu}$ simultaneously.

\paragraph{Cosmological decomposition.---}
It was argued previously that, in general, the decomposition can be characterized with two geometrical objects compatible with the symmetry, $T_\mu = (1,\vec{0})$ and a homogeneous and isotropic spatial metric $s_{ij}$\footnote{The analysis presented in this section is not valid in three dimensions, due to the existence of the skew-symetric tensor \(\epsilon_{tij}\). The lower dimensional cases will be analyzed in independently.} as follows
\begin{dmath}
  Y_{\lambda\mu\nu}
  =\left(P_{\lambda\mu\nu}^{\lambda'\mu'\nu'}-\widetilde{P}_{\lambda\mu\nu}^{\lambda'\mu'\nu'}\right) B \, T_{\lambda'}\delta_{\mu'}^i \delta^j_{\nu'}s_{ij}
  =0,
\end{dmath}  
i.e. there are no homogeneous and isotropic contributions to the connection with the tensor symmetries of $Y_{\lambda\mu\nu}$.

\paragraph{Static spherical decomposition.---}
In this case the decomposition is characterized by three geometrical objects, $T_\mu = (1,0,\vec 0)$, $R_\mu = (0,1,\vec 0)$  and $s_{ab}$ the metric of the $n-2$ dimensional sphere the only objects symmetric under time displacements and rotations, we also have reflections and time reversal symmetry, so, the most general such tensor is
\begin{dmath}
  Y_{\lambda\mu\nu}
  =\left(P_{\lambda\mu\nu}^{\lambda'\mu'\nu'}-\widetilde{P}_{\lambda\mu\nu}^{\lambda'\mu'\nu'}\right) \left(A \, R_{\lambda'} T_{\mu'} T_{\nu'} + B \, R_{\lambda'} \delta_{\mu'}^a \delta^b_{\nu'} s_{ab}\right)
  = Y(r)\left((n-2)F(R_{\lambda}T_{\mu}T_{\nu}-T_{\lambda}T_{(\mu}R_{\nu)})
    + r^2 (R_{\lambda}\delta_{\mu}^a \delta^b_{\nu} 
    - R_{(\mu} \delta^a_{\nu)}\delta_{\lambda}^b) s_{ab}\right),
\end{dmath} 
which leaves a single degree of freedom.

\subsection{Decomposition of the vectors}

As the trace of nonmetricity, these should be considered tensors under coordinate transformations, yet, the independence of $S_{\lambda\mu\nu}$ on the conformal factor of the metric, leaves a gauge symmetry, which we are going to deal with once we try to equate geodesic and autoparallel equations. 
For vectors and covectors the thing is pretty straightforward in cosmological and static black hole scenarios
\begin{equation}
  \begin{aligned}
    V^\mu & = V(t) \, T^\mu,\\
    W_\mu & = W(t) \, T_\mu,\\
    A_\mu & = A(t) \, T_\mu,
  \end{aligned}
\end{equation}
where the index of $T^\mu$, the only vector symmetric under the cosmological principle, was raised by the metric.

For static spherical symmetry, with time reversal invariance, the decomposition of the vectors is,
\begin{equation}
  \begin{aligned}
    V^\mu & = V(r) \, R^\mu,\\
    W_\mu & = W(r) \, R_\mu,\\
    A_\mu & = A(r) \, R_\mu.
  \end{aligned}
\end{equation}

\subsection{Trace-less part of the torsion}

The tensor decomposition of the traceless torsion could be used to separate transverse from longitudinal degrees of freedom, as
\begin{equation}
  [B_{\lambda\mu\nu}] - [B^{TT}_{\lambda\mu\nu}]
  = [\sigma_{\mu\nu}] + [\omega^T_{\mu\nu}]
\end{equation}
where $\sigma_{\mu\nu}$ is a symmetric tensor whose decomposition we are all familiar with, and $\omega^T_{\mu\nu}$ is a transverse antisymmetric tensor.

\paragraph{Cosmological decomposition.---} Following a similar algorithm as before, one can naively conclude that there are no nontrivial cosmological components coming from the traceless part of the torsion, i.e,
\begin{equation}
  \label{eq:cosmo-B-higer-dimen}
  B_{\lambda\mu\nu} = 0.
\end{equation}
However, in four dimensions, Eq.~\eqref{eq:cosmo-B-higer-dimen} has to be modified due to the existence of the invariant spatial tensor \(\epsilon_{ijk}\). Hence, in four dimensions the decomposition is 
\begin{equation}
  \label{eq:cosmo-B-4d}
  B_{\lambda\mu\nu} = B(t) \epsilon_{ijk} \, \delta^i_{\lambda} \delta^j_{\mu} \delta^k_{\nu}.
\end{equation}

\paragraph{Static spherical decomposition.---} Similarly, the static spherical decomposition of this traceless tensor, compatible with the time-reversal condition, gets
\begin{equation}
  \label{eq:sphere-B-hd}
  B_{\lambda\mu\nu} = B(r) \left((n-2) F \, T_{\lambda}T_{[\mu}R_{\nu]}
    + r^2 s_{ab} \delta_\lambda^a \delta_{[\mu}^b R_{\nu]} \right),
\end{equation}
although, in four dimensions, other terms are possible and the traceless torsion is
\begin{dmath}
  \label{eq:sphere-B-4d}
  B_{\lambda\mu\nu} = B(r) \left(2 F \, T_{\lambda}T_{[\mu}R_{\nu]}
    + r^2 s_{ab} \delta_\lambda^a \delta_{[\mu}^b R_{\nu]} \right)
    +C(r) \sqrt{s}\epsilon_{ab} \delta_\lambda^a \delta_{[\mu}^b R_{\nu]}
    +D(r) \sqrt{s}\epsilon_{ab} R_{\lambda}\delta_\mu^a \delta_{\nu}^b ,
\end{dmath}where $\epsilon_{ab}$ is the Levi-Civita tensor on the sphere.

In what follows we are going to assume that the ansatz for the connection with certain symmetries, such as for cosmological solutions or black holes, can be represented with the connections of the subspaces, which leads to the study of the decomposition of the connections in lower dimensional models. Thus, we are going to study connections in dimensions two and three, and their symmetry reductions.

\subsection{Geodesics}
The affine connection for the cosmological and static black hole scenarios is simplified by
\begin{dmath}
  \label{nd}
  \frac{dU^\mu}{d\tau}
  + \Gamma_{\lambda}{}^{\mu}{}_{\kappa}(g) U^\lambda U^\kappa
  + U^\mu W_\lambda U^\lambda
  + U^2 V^\mu
  + Y^\mu{}_{\lambda\kappa} U^\lambda U^\kappa
  =
  0.
\end{dmath} This is because $S^{TT}_{\lambda\mu\nu}$ can be set to zero as a gauge choice, although we determined that there may be some residual gauge after the condition that it is transverse, no additional degrees of freedom are left in it, and a nonzero choice would only modify the relations between the overdetermined connection in metric affine variables vs the raw decomposition of the symmetric affine connection. 

\subsubsection{Geodesics in cosmology}

In the cosmological scenario, the geodesic equation is
\begin{dmath}
  \frac{dU^\mu}{d\tau}
  + \Gamma_{\lambda}{}^{\mu}{}_{\kappa}(g) U^\lambda U^\kappa
  + U^\mu WT_\lambda U^\lambda
  + U^2 VT^\mu
  =
  0,
\end{dmath}
and with the additional condition $V=0$, the geodesics of $g_{\mu\nu}$ can be identified with autoparallels.

\subsubsection{Geodesics in static spherical solutions}

The geodesic equation in the static spherical scenario is
\begin{dmath}
  \frac{dU^\mu}{d\tau}
  + \Gamma_{\lambda}{}^{\mu}{}_{\kappa}(g) U^\lambda U^\kappa
  + Y(r)\left((n-2)F(R^{\mu}T_{\lambda}T_{\kappa}-T^{\mu}T_{(\lambda}R_{\kappa)})
    +  (r^2R^{\mu}\delta_{\lambda}^a \delta^b_{\kappa} s_{ab}
    - R_{(\lambda} \delta^a_{\kappa)}\delta_a^{\mu}) \right)U^\lambda U^\kappa
  + U^\mu WR_\lambda U^\lambda
  + U^2 VR^\mu
  =
  0.
\end{dmath}
Now, unless $Y=0$ and $V=0$, there is no way in which autoparallels can be identified with geodesics, but this would seriously limit the space of connections we would be studying. 
Instead, we are going to explore the behavior of radial autoparallels, those whose angular components are zero ($U^a=0$), since the identification with radial geodesics may open up the field of black hole physics within affine geometry. Hence,
\begin{dmath}
  \label{tnd}
  \frac{dU^t}{d\tau}
  + \Gamma_{\lambda}{}^{t}{}_{\kappa}(g) U^\lambda U^\kappa
  + (W+(n-2)Y) U^t U^r
  =
  0.
\end{dmath}
\begin{dmath}
  \frac{dU^r}{d\tau}
  + \Gamma_{\lambda}{}^{r}{}_{\kappa}(g) U^\lambda U^\kappa
  + (W +(n-2)Y)(U^r)^2
  + U^2 (V- (n-2)Y)\frac{F}{G}
  =
  0.
\end{dmath}
Thus, with the additional condition $V=(n-2)Y$, the geodesics of $g_{\mu\nu}$ can be identified with autoparallels.

\section{Connections in two dimensions\label{sec:conn-two-dimens}}

The decomposition of the connection shown in Eq.~\eqref{ConnectionDecomposition} is not valid in two dimensions, because the traceless tensor with mixed symmetry \(Y^\lambda{}_{\mu\nu}\) vanishes,\footnote{Note that it has $\frac{1 \cdot 2 \cdot 3}{1 \cdot 3 \cdot 1} - 2 = 0$ independent components.} and the Helmholtz decomposition in terms of transverse and longitudinal components of the symmetric tensor $S_{\lambda\mu\nu}$ leads us to conclude that the transverse part has no independent components as well, for which we can safely assume that $S_{\lambda\mu\nu}$ is the traceless part of a tensor of the form $\nabla_{(\lambda}\sigma_{\mu\nu)}$ with no transverse component. This immediately leads us to the conclusion that, through a gauge transformation of the metric [Eq.~\eqref{metricGeneralTransformations}], the connection simplifies to 
\begin{equation}
  \Gamma_{\mu}{}^{\lambda}{}_{\nu} = \Gamma_{\mu}{}^{\lambda}{}_{\nu}(g) + V^\lambda g_{\mu\nu} + 2 W_{(\mu} \delta^\lambda_{\nu)}.
\end{equation}
The six components of the connection can be described by the three components of the metric together with the two components of $V^\mu$ and the two components of $W_\mu$. There appears to be one extra component when one compares the ones of the connection with those of the tensors,
but the trace of the metric compatible connection  $\Gamma_{\mu}{}^{\nu}{}_{\nu} (g) = \partial_\mu \ln{\sqrt{g}}$ can be reabsorbed by a gauge transformation of the fields $V^\mu$ and $W_\mu$.

The role of the nonmetricity in two dimensions is related to transformations we can perform in the geometry. We can see this clearer by taking the covariant derivative of the metric
\begin{dmath}
  \nabla^\Gamma_\lambda g_{\mu\nu}
  =
  \nabla^g_\lambda g_{\mu\nu}
  - 2 V^\kappa g_{\lambda(\mu}g_{\nu)\kappa}
  - 2 W_{(\lambda} \delta^\kappa_{\mu)} g_{\kappa\nu}
  - 2 W_{(\lambda} \delta^\kappa_{\nu)} g_{\mu\kappa}
  =
  - 2 W_\lambda g_{\mu\nu}
  - 2 g_{\lambda(\mu} (W_{\nu)}+V_{\nu)}), 
\end{dmath}
note that if $W_\mu = - V_\mu$, the combined role of these vectors is well known in the literature as  the Weyl connection, used to achieve invariance under conformal transformations, i.e. $g_{\mu\nu} \to g^{\prime}_{\mu\nu} = e^{2 \phi} g_{\mu\nu}.$ 

We can understand the geometric meaning of $W_\mu$ if we observe its role in the parallel transport of tensors, say $U^\mu \nabla^\Gamma_\mu T^{\cdots}{}_\cdots$, and more specifically, the role of \(W_\mu\) on autoparallel curves 
\begin{equation}
  \label{autoparallelEquations}
  \frac{DU^\mu}{D\tau} = U^\lambda \nabla^\Gamma_\lambda U^\mu = 0,
\end{equation}
defined by its tangent vector $U^\mu=\frac{dx^\mu}{d\tau}$, with a given parametrization of the curve $\tau$.

Hence,
\begin{dmath}
  \frac{dU^\mu}{d\tau}
  + \Gamma_{\lambda}{}^{\mu}{}_{\kappa}(g) U^\lambda U^\kappa
  + U^\mu W_\lambda U^\lambda
  + U^2 V^\mu
  =
  0,
\end{dmath}
where  $U^2=U^\lambda U^\kappa g_{\lambda\kappa}$ is the squared norm of the curve's velocity.

Geodesics, i.e. curves that minimize the metric distance between two points, are defined by the metric connection, however, a different choice of the curve's parameter such that
\begin{equation}
  \tau \to \tau = f(\tau'),
\end{equation}
will introduce the term  $\frac{d^2\tau'}{d\tau^2}U^\mu$, which can be used to cancel the term with $W_\lambda$, and if $V^\mu=0$ autoparallels would be indistinguishable from geodesics. 

An interesting particular case of study is when $V_\mu = \partial_\mu\nu$. In this case, we can perform a conformal transformation, yielding
\begin{dmath}
  \Gamma_{\mu}{}^{\lambda}{}_{\nu}
  =
  \Gamma_{\mu}{}^{\lambda}{}_{\nu}(g')
  + \Big(V_\kappa-\partial_\kappa\phi\Big) g^{\prime\lambda\kappa} g'_{\mu\nu}
  + 2 \Big(W_{(\mu} + \partial_{(\mu} \phi\Big) \delta_{\nu)}^\lambda.
\end{dmath}
If we choose $\phi=\nu$ we get $V'_\mu=0$. Hence, all autoparallels become geodesics. This argument can be employed in the cosmological and static spherical scenarios, but can also be used in a more general context. Otherwise, one can get rid of the longitudinal pat of \(V\) and keep the transverse vector $V^\mu=\frac{1}{\sqrt{g}}\epsilon^{\mu\nu}\partial_\nu v$.

Moreover, the term $U^2 V^\mu$, when non-vanishing, plays the role of a geodesic deviation's force, which, for null-geodesics has no contribution, i.e. null geodesics are indistinguishable from autoparallels. The evolution of the squared norm can also be computed, and we get
\begin{equation}
  \label{2dNormEvolution}
  \frac{DU^2}{D\tau}
  = U^\lambda U^\mu U^\nu \nabla^\Gamma_\lambda g_{\mu\nu}
  = - 2 U^2 U^\mu (2 W_\mu + V_\mu),
\end{equation}
which, for $U^2 = 0$, implies the conservation of the norm along autoparallel curves.

For the conservation of non-zero norms we have to set $V_\mu = - 2 W_\mu$, or choose a specific parametrization of the curve that allows to eliminate the whole term, key for this procedure is that the term is proportional to $U^2$ and so does the term that appears when we reparametrize $\tau$.

\subsection{Cosmological decomposition}

An ansatz for the two-dimensional connection that follows the cosmological principle (homogeneous and isotropic) can be implemented easily, by choosing $(x^0, x^1) = (t, x)$ we set
\begin{equation}
  V^\mu=(V(t),0), \quad W_\mu=(W(t),0),
\end{equation}
and
\begin{equation}
  g_{\mu\nu} = \diag(-N^2(t),\ a^2(t)).
\end{equation}

A na\"ive naming of the non-zero components of the connection  can now be compared to the choice we have proposed in terms of the metric, and the conformal and projective vectors, since
\begin{align}
  \Gamma_{0}{}^{0}{}_{0} & = J = \frac{\dot N}{N}-N^2V+2W,\\
  \Gamma_{1}{}^{0}{}_{1} & = g = \frac{a\dot a}{N^2}+a^2V,\text{ and}\\
  \Gamma_{0}{}^{1}{}_{1} & = h = \frac{\dot a}{a}+W.
\end{align}
Given $(N,a,V,W)$, we can easily find $(g,h,J)$. For the reverse, i.e. to determine $(N,a,V,W)$ from $(g,h,J)$, we are missing a condition.

This mismatch can be compensated by selecting a relation among the variables. For example, if we would like to have a parameter of the curve such that the linear momentum can be defined by $p^\mu=mU^\mu$ and $p_\mu p^\mu$ is a conserved quantity along the path followed by a freely falling particle, then, in accordance with Eq.~\eqref{2dNormEvolution}, we may choose $V=2W/N^2$. We could instead choose one of the many other arbitrary conditions $F(N,a,V,W)=0$ that does not limit the possible values of  $(g,h,J)$. For instance, we could use $N=1$ as a convenient gauge choice (of an otherwise arbitrary metric) in order to compare solutions with Friedmann--Robertson--Walker models of cosmology.

A specially reasonable choice is to set $V=0$, where all autoparallel curves are equivalent to geodesics, and find the metric whose geodesics are the autoparallels of any arbitrary cosmological connection. The question we should face is whether or not a cosmological metric and a projective vector \(W_{\mu}\) are enough to reproduce any cosmological metric. The system of equations for $(N,a,W)$ 
\begin{align}
  \label{eq:cosmo-relations-2d}
  g & = \frac{a\dot a}{N^2}, &
  h & = \frac{\dot a}{a}+W, &
  J & = \frac{\dot N}{N}+2W,
\end{align}
can be used to get
\begin{equation}
  2 h - J = 2 \frac{\dot a}{a} - \frac{\dot N}{N},
\end{equation}
whose  solution is
\begin{equation}
  \label{eq:cosmo-solution-2d}
  \frac{a^2}{N} = \frac{a^2_0}{N_0} \exp\left\{\int_0^t dt'(2h - J) \right\}.
\end{equation}
From Eqs.~\eqref{eq:cosmo-relations-2d} and~\eqref{eq:cosmo-solution-2d} we also get 
\begin{equation}
  g\Big(\frac{a^2}{N}\Big)^{-2}=\frac{\dot a}{a^3},
\end{equation}
from which obtain
\begin{equation}
  a^2 = \frac{a_0^2}{1 - 2 \frac{N^2_0}{a^2_0} \int_0^{t} dt^{\prime} g(t^{\prime}) \exp\left\{-2\int_0^{t^{\prime}} dt^{\prime\prime} ( 2 h - J ) \right\} },
\end{equation}
that can be used, together with the previous solution, to obtain $N$. Finally, we obtain $W$ using Eq.~\eqref{eq:cosmo-relations-2d}.

From these expressions for $(a,N,W)$ in terms of $(g,h,J)$, we infer that the change of variables in the kinematics of the connection is invertible and that it limits the space of solutions in no way.

\subsection{Static spherical ansatze}

Consider now a static spherically symmetric spacetime with time reversal symmetry with coordinates
\begin{equation}
  (x^0, x^1) = (t, r).
\end{equation}
In this framework the ansatz would traditionally look like
\begin{align}
  V^\mu & = (0,V(r)), &
  W_\mu & = (0,W(r)),
\end{align}
and 
\begin{equation}
  g_{\mu\nu} = \diag(-F(r),\ G(r)/F(r)).
\end{equation}

A substitution of $(r,t) \leftrightarrow (t,x)$ takes us from one of the models to the other and all things said in the previous section hold. 

In general, the kinematics of the black hole scenario is bigger than just the static or even the stationary configurations, see for example Ref.~\cite{castillo-felisola24_polyn_affin_model_gravit}. Traditionally, the Birkhoff theorem in more dimensions saves us from the study of dynamic spherically symmetric solutions, but that occurs in the context of metric models, we expect this can be clarified in the context of polynomial affine model of gravity as well, whether it is possible to establish a Birkhoff-like theorem or not.  

In higher dimensional polynomial affine models of black holes,  it may prove helpful to restrict to radial autoparallels, once we have found the two-dimensional metric whose geodesics are autoparallel, we could find the trapped regions for timelike geodesics. The extension to higher dimensions may require to define an extension of the two-dimensional restriction of the metric such that trapped non-radial autoparallels are defined as timelike too.

\section{Connections in three dimensions\label{sec:conn-three-dimens}}

In two dimensions we saw that the na\"ive ansatz and the ansatz in terms of a metric and nonmetricity are equivalent. In three dimension though, without assuming special symmetries, we can just assume all eighteen components of the connection can be represented through an irreducible tensor decomposition.

After a choice of a generic metric, we have that
\begin{equation}
  \Gamma_{\mu}{}^{\lambda}{}_{\nu}
  =
  \Gamma_{\mu}{}^{\lambda}{}_{\nu}(g)
  + S{}^\lambda{}_{\mu\nu}
  + Y{}^\lambda{}_{\mu\nu}
  + V^\lambda g_{\mu\nu}
  + 2 W_{(\mu}\delta^\lambda_{\nu)},
\end{equation}
has eighteen components on the left-hand side, while on the right-hand side (counting from left to right) appear to have twenty-four components \((6 + 7 + 5 + 3 + 3)\), which has an excess of six components. In fact, the seven components of the symmetric tensor $S_{\lambda\mu\nu}$, can be decomposed into transverse and non transverse. In the transverse part, all indices are transverse, and they behave as if they belong in a \((d-1)\)-dimensional space, thus, the fully symmetric traceless and transverse \(3\)-tensor has $\frac{4 \cdot 3 \cdot 2}{3 \cdot 2} - 2 = 2$ independent components, while the symmetric contributions to the non-transverse symmetric part of the connection is represented by the metric connection. The degrees of freedom counting from left to right becomes \(6 + 2 + 5 + 3 + 3 = 19\), but the trace  of the connection can also be redefined through conformal transformations and absorbed as a gauge transformation of the conformal vector field. Notice that the choice of metric is completely arbitrary (a gauge choice) but it allows us to perform the tensor decomposition. Yet, we can use them as kinematic degrees of freedom by setting some constraints on the components of the other tensors---in this case, that $S_{\lambda\mu\nu}$ is transverse.

Following the steps taken to study the two-dimensional connection, the covariant derivative of the metric reveals the nonmetricity as
\begin{dmath}
  \nabla^\Gamma_\lambda g_{\mu\nu}
  =
  \nabla^g_\lambda g_{\mu\nu}
  - 2 V^\kappa g_{\lambda(\mu} g_{\nu)\kappa}
  - 2 W_{(\lambda} \delta^\kappa_{\mu)} g_{\kappa\nu}
  - 2 W_{(\lambda} \delta^\kappa_{\nu)} g_{\mu\kappa}
  - 2 S^\kappa{}_{\lambda(\mu} g_{\nu)\kappa}
  - 2 Y^\kappa{}_{\lambda(\mu} g_{\nu)\kappa} 
  =
  - 2 W_\lambda g_{\mu\nu}
  - 2 g_{\lambda(\mu} W_{\nu)}
  - 2 g_{\lambda(\mu} V_{\nu)}
  - 2 S_{\lambda\mu\nu}
  - Y_{\lambda\mu\nu}.
\end{dmath}
Autoparallels are defined by the Eq.~\eqref{autoparallelEquations}, or more explicitly
\begin{dmath}
  \frac{dU^\mu}{d\tau}
  + \Gamma_{\lambda}{}^{\mu}{}_{\kappa}(g) U^\lambda U^\kappa
  + U^\mu W_\lambda U^\lambda
  + U^2V^\mu
  + S^\mu{}_{\lambda\kappa} U^\lambda U^\kappa
  + Y^\mu{}_{\lambda\kappa} U^\lambda U^\kappa
  =
  0,
\end{dmath}
and the squared norm evolution by 
\begin{dmath}
  \label{NormEvolution}
  \frac{DU^2}{D\tau}
  =
  U^\lambda U^\mu U^\nu \nabla^\Gamma_\lambda g_{\mu\nu}
  =
  - 2 U^2 U^\mu (2W_\mu + V_\mu)
  - 2 S_{\lambda\mu\nu} U^\lambda U^\mu U^\nu,
\end{dmath} 
where $S_{\lambda\mu\nu}$ can be safely assumed to be traceless and transverse, but still nonzero in the most general case. The last term implies that the norm of a vector cannot be conserved over autoparallels unless certain symmetries apply. 

\subsection{Cosmological ansatz}

Isotropy and homogeneity can be imposed on solutions. In order to represent the split between the dimensions of the homogeneous space and time, we use greek letters for the full space and latin letters from the beginning of the alphabet such that $x^\mu \to (t,x^a)$.

We propose the cosmological metric
\begin{equation}
  g_{\mu\nu} = \diag(-N^2, a^2 s_{ab}),
\end{equation}
where $s_{ab} = \diag((1-\kappa r^2)^{-1}, r^2)$, with $\kappa = - 1, 0, 1$. We can represent $Y^\lambda{}_{\mu\nu}$ using the covariant Levi-Civita skew-symmetric tensor $\tilde{\epsilon}_{ab} = \sqrt{s} \epsilon_{ab}$ as
\begin{dmath}
  Y_{\lambda\mu\nu}
  =Y(r)2 \tilde\epsilon_{ab}\delta^a_{\lambda}\delta_{(\mu}^bT_{\nu)}.
\end{dmath} 
Finally, we also evaluate the traceless torsion that can be expressed in terms of two functions
\begin{dmath}
    B_{\lambda\mu\nu}=B(r) \sqrt{s}\epsilon_{ab} \delta_\lambda^a \delta_{[\mu}^b T_{\nu]}
    +C(r) \sqrt{s}\epsilon_{ab} T_{\lambda}\delta_\mu^a \delta_{\nu}^b.
\end{dmath}
The split of the connection reveals that the only nonzero components of the connection are
\begin{align}
  \Gamma_{0}{}^{0}{}_{0} & = J = \frac{\dot N}{N} - N^2 V + 2 W,\\
  \Gamma_{a}{}^{0}{}_{b} & = g s_{ab} = \Big(\frac{a\dot a}{N^2}+a^2V\Big) s_{ab},\\
  \Gamma_{0}{}^{a}{}_{b} & = h \delta^a_b + f \tilde\epsilon^a{}_b = \Big(\frac{\dot a}{a}+W\Big) \delta^a_b + Y \tilde\epsilon{}^a{}_b,\\
  \Gamma_{a}{}^{c}{}_{b} & = \gamma_{a}{}^{c}{}_{b}(s),
\end{align}
where $\gamma_{a}{}^{c}{}_{b}(s)$ is the connection compatible with the metric \(s\), i.e. $\nabla^{\gamma}_c s_{ab}=0$.

In three dimensions the na\"{\i}ve ansatz for the connection includes a term that cannot be included in any other dimensionality, but it is fairly clear that it comes from $Y^a{}_{0b}$, and there is a clear identification $f = Y$, while the other variables follow the same arguments from the two-dimensional case. We conclude that the connection can be described successfully through the set of variables $(N,a,V,W,Y)$, although the additional condition $V=0$ which, for cosmological solutions with $Y=0$, allows for the identification of geodesics and autoparallels. This condition is very restrictive, though, but the geodesic equation in components
\begin{dmath}
  \frac{dU^t}{d\tau}
  + \Gamma_{\lambda}{}^{t}{}_{\kappa}(g) U^\lambda U^\kappa
  + U^t W U^t
  + U^2 V
  =
  0,
\end{dmath}
and
\begin{dmath}
  \frac{dU^a}{d\tau}
  + \Gamma_{\lambda}{}^{a}{}_{\kappa}(g) U^\lambda U^\kappa
  + U^a W U^t
  + 2Ys^{ab}\sqrt{s}\epsilon_{bc} U^c U^t
  =
  0,
\end{dmath}
leaves no alternative since the $Y$ term cannot be reabsorbed.

\subsection{Black Hole ansatz}

A black hole solution is generally assumed to have an axial symmetry, which in $2+1$ dimensions is translated to having rotational symmetry. This symmetry does not imply parity invariance, so, for instance, when flipping the angular direction, a stationary rotating BH rotates in the opposite direction.
We shall describe the kinematics of stationary black holes using the metric and nonmetricity decomposition, but we shall see a remarkable reduction in degrees of freedom after imposing that the solutions are symmetric under time reversion and angular parity to describe the kinematics of static black holes.

Yet, there are some details about the definitions, black holes are defined by the presence of a curvature singularity and the presence of a null surface reachable in finite time that surrounds the singularity. In order to define what we mean by black hole in non-Riemannian geometries, we will consider three cases: (1) the ansatz for time dependent black hole connections; (2) the stationary black hole connections; and finally (3) the static black hole connections. We could keep the discussions separate even so there is not much difference on some aspects. Then we shall explore the geodesics and invariant tensors to evaluate possible means to determine when a solution is a black hole.

\subsubsection{Ansatze}

From the general decomposition of the connection
\begin{equation}
  \Gamma_{\mu}{}^{\lambda}{}_{\nu}
  =
  \Gamma_{\mu}{}^{\lambda}{}_{\nu}(g)
  + Y{}^\lambda{}_{\mu\nu}
  + V^\lambda g_{\mu\nu}
  + 2 W_{(\mu} \delta^\lambda_{\nu)},
\end{equation}
with 
\begin{equation}
  Y^\lambda{}_{\mu\nu}
  =
  Y_{\kappa(\mu} \tilde{\epsilon}_{\nu)}{}^{\kappa\lambda},
\end{equation}
and where we have set the fully symmetric, traceless and transverse $S_{\lambda\mu\nu} = 0$.

\paragraph{Time Dependent Black Hole Connections.---}
The black hole solution subspace of connections has rotational symmetry, which in two spatial dimensions is generated by a constant vector field $X = \partial_\theta$ and $\mathcal{L}_X \Gamma_{\mu}{}^{\lambda}{}_{\nu} = 0$ imposes the independence of the connection on $\theta$, but no further restrictions are required on the components of the field.

Using coordinates $(t,r,\theta)$ we can propose a split of the tensor components in $\mu=(a,\theta)$, which will allow us to establish a useful naming convention when we further restrict the model. The resulting quantities are then,
\begin{align}
  g_{\mu\nu}
  & =
    \begin{pmatrix}
      q_{ab} & p_b\\
      p_a & R^2
    \end{pmatrix},
  \\
  Y_{\mu\nu}
  & = 
    \begin{pmatrix}
      Z q_{ab} + X_{ab} & Y_a\\
      Y_b & - 2 Z R^2
    \end{pmatrix},
  \\
  V^{\lambda}
  & =
    \begin{pmatrix}
      V^a & V^\theta
    \end{pmatrix},
  \\
  W_{\lambda}
  & =
    \begin{pmatrix}
      W_a & W_\theta
    \end{pmatrix},
\end{align}
where all the objects depend on the coordinates \(t\) and \(r\), and also \(X_{[ab]} = 0\) and hence \(X_{ab} q^{ab} = 0\).

\paragraph{Stationary Black Hole Connections.---}
Since flipping the time direction and flipping the rotation axis will both have the consequence of making the black hole look as reversing its rotation direction we can choose coordinates where the connection's symmetry under $(t,\theta) \to (-t,-\theta)$ is explicit,
\begin{align}
  g_{\mu\nu}
  & =
    \begin{pmatrix}
      q_{ab}(r)&p(r)\delta^t_b\\
      p(r)\delta^t_a&r^2
    \end{pmatrix},
  \\
  q_{ab}
  & =
    \begin{pmatrix}
      -F(r)&0\\
      0&\frac{G(r)}{F(r)}
    \end{pmatrix},
  \\
  Y_{\mu\nu}
  & = 
    \begin{pmatrix}
      Z(r)q_{ab}(r)+X_{ab}(r) & Y_a(r)\\
      Y_b(r)& -2Z(r)r^2
    \end{pmatrix},
  \\
  V^{\lambda}
  & =
    \begin{pmatrix}
      V^a(r) & 0
    \end{pmatrix},
  \\
  W_{\lambda}
  & =
    \begin{pmatrix}
      W_a(r) & 0
    \end{pmatrix}.
\end{align}

\paragraph{Static Black Hole Connections.---}
We can  represent $Y^\lambda{}_{\mu\nu}$ as
\begin{dmath}
  Y_{\lambda\mu\nu}
  =Y(r)\left(F(R_{\lambda}T_{\mu}T_{\nu}-T_{\lambda}R_{(\mu}T_{\nu)})
    +r^2(R_{\lambda}\delta_{\mu}^\theta \delta^\theta_{\nu}-\delta_{\lambda}^\theta \delta^\theta_{(\mu}R_{\nu)})\right),
\end{dmath} 
\begin{align}
  g_{\mu\nu}
  & =
    \begin{pmatrix}
      q_{ab}(r)&0\\
      0&r^2
    \end{pmatrix},
  \\
  q_{ab}
  & = 
    \begin{pmatrix}
      -F(r)&0\\
      0&\frac{G(r)}{F(r)}
    \end{pmatrix},
  \\
  V^{\lambda}
  & =
  \begin{pmatrix}
    V(r)\delta_r^a & 0
  \end{pmatrix}
  = V(r)\delta_r^\lambda,
  \\
  W_{\lambda}
  & =
    \begin{pmatrix}
      W(r)\delta^r_a & 0
    \end{pmatrix}
    = W(r)\delta^r_\lambda.
\end{align}
In this case, we also point out that the traceless torsion can be expressed as
\begin{dmath}
    B_{\lambda\mu\nu}=B(r) \left( F \, T_{\lambda}T_{[\mu}R_{\nu]}
    + r^2 \delta_\lambda^\theta \delta_{[\mu}^\theta R_{\nu]} \right)
\end{dmath}

\subsubsection{Geodesics}

From the general geodesic with $S_{\lambda\mu\nu} = 0$, we get the following results.

\paragraph{Time Dependent Black Hole Connections.---}
For this case, the geodesic equation is 
\begin{dmath}
  \frac{dU^\mu}{d\tau}
  + \Gamma_{\lambda}{}^{\mu}{}_{\kappa}(g) U^\lambda U^\kappa
  + U^\mu W_\lambda U^\lambda
  + U^2 V^\mu
  + Y^\mu{}_{\lambda\kappa} U^\lambda U^\kappa
  =
  0,
\end{dmath}
and the presence of $Y^\mu{}_{\lambda\kappa}U^\lambda U^\kappa$ makes it improbable to be able to choose  a metric whose geodesics coincide with autoparallels.
It is possible to concentrate our efforts in studying radial geodesics, 

\paragraph{Static Black Hole Connections.---}
Geodesics for the static black hole solutions are further simplified, 
\begin{dmath}
  \label{SBH3d}
  \frac{dU^\mu}{d\tau}
  + \Gamma_{\lambda}{}^{\mu}{}_{\kappa}(g) U^\lambda U^\kappa
  + U^\mu W_\lambda U^\lambda
  + U^2 V^\mu
  + Y^\mu{}_{\lambda\kappa} U^\lambda U^\kappa
  =
  0,
\end{dmath}
where 
\begin{dmath}
  Y^t{}_{\lambda\kappa} U^\lambda U^\kappa = Y(r)U^tU^r,
\end{dmath}
\begin{dmath}
  Y^r{}_{\lambda\kappa}U^\lambda U^\kappa
  =
  Y(r) \frac{F}{G} (F(U^t)^2 + r^2(U^\theta)^2),
\end{dmath}
and
\begin{dmath}
  Y^\theta{}_{\lambda\kappa}U^\lambda U^\kappa
  =
  - Y(r) U^\theta U^r.
\end{dmath}
Of particular interest are radial autoparallels ( $U^\theta=0$), for which the norm defined by the proposed metric is
\begin{equation}
  U^2 = - F (U^t)^2 + \frac{G}{F} (U^r)^2.
\end{equation}
A separation of the metric connection components  in this subspace leads us to $\Gamma_{a}{}^{c}{}_{b}(g) = \Gamma_{a}{}^{c}{}_{b}(q)$.
This allows us to rewrite (\ref{SBH3d}) as
\begin{dmath}
  \label{SBH3d'}
  \frac{dU^\mu}{d\tau}
  + \Gamma_{\lambda}{}^{\mu}{}_{\kappa}(q) U^\lambda U^\kappa
  + U^\mu \left(W_\lambda + \delta^r_\lambda Y(r)\right) U^\lambda
  + U^2 \left(V^\mu - \delta^\mu_r\frac{F}{G}Y(r)\right)
  =
  0.
\end{dmath}
With this, in order to identify autoparallels with radial geodesics, we can perform an $r$ dependent conformal transformation $g'_{\mu\nu} = e^{2\varphi(r)} g_{\mu\nu}$  to set $V^\mu = \delta^\mu_r \frac{F}{G}Y$. This gauge choice of the metric allows for the last term in Eq.~\eqref{SBH3d'} to be zero and only an affine reparametrization of the curve would be necessary to identify the autoparallels equation to the one for geodesics. 

\subsubsection{Norms}

Although we saw in the last section that geodesics and autoparallels cannot be conciliated except in the case of static black hole solutions,  there is a special type of geodesics that is preserved  when $S_{\lambda\mu\nu}=0$, null geodesics $U^2=0$. This occurs because 
\begin{dmath}
  \frac{DU^2}{D\tau}
  =
  U^\lambda U^\mu U^\nu \nabla^\Gamma_\lambda g_{\mu\nu}
  =
  - 2 U^2 U^\mu (2W_\mu + V_\mu)
  + S_{\lambda\mu\nu} U^\lambda U^\mu U^\nu,
\end{dmath} 
and setting the transverse and traceless $S_{\lambda\mu\nu}$ to zero only leaves terms that can be reabsorved in the geodesic equation by a specific choice of the curve's parameter.

\section{Concluding remarks\label{sec:concluding-remarks}}

In the context of modified and extended models of gravity, one commonly encounters three types of theories, depending on whether their supporting geometry is metric, affine, or metric-affine. The main difference between these geometries lies in the fundamental objects that support the geometry.

Assuming that the dimension of the manifold is \(n\), a simple counting of independent components reveals that metric geometries are characterized by the \(n(n+1)/2\) components of the metric, affine geometries by the \(n^3\) components of the affine connection, and metric-affine geometries by the combined \(n(2 n^2 + n + 1)/2\) of the metric and affine connection. From these cases, the affine models of gravity have been consigned to an \emph{inferior} category, and consequently have been less investigated, probably for the lack of an intuitive interpretation of their fundamental object, the affine connection.

In order to overcome the difficulty of dealing with affine models of gravity, we have proposed a decomposition of the connection, using an \emph{auxiliary} metric tensor field. Unlike the metric-affine case, the \emph{auxiliary} metric does not introduce new degrees of freedom, it just allows to choose a reparametrization of the connection. Utilizing this auxiliary metric, one can decompose the affine connection into three components, to know the Levi-Civita component, the contorsion and the deflection. This decomposition has been known since the earliest days of the tensor calculus; see, for example, Refs.~\cite{eisenhart27_non_rieman,weyl22_space,schouten13_ricci}. However, considering the metric tensor as an auxiliary field implies that it doesn't take a leading role in constructing the model.

From the above discussion, it should be clear that alternative selections of the auxiliary metric do not affect the supporting geometry. Hence, the choice of metric is a gauge redundancy.

We have shown how the components of the connection transform under (infinitesimal) transformations of the auxiliary metric
\begin{equation*}
  g_{\mu\nu} \to g^{\prime}_{\mu\nu} = g_{\mu\nu} + s_{\mu\nu}.
\end{equation*}
Although changing the metric modifies the decomposition of the connection, e.g. its Levi-Civita component, our results confirm that the affine geometry remains the same.

The introduction of the auxiliary metric allows us to decompose the (symmetric) affine connection in up to five parts: \(\Gamma(g)\), \(S\), \(Y\), \(V\) and \(W\). However, the explicit decomposition of the connection differs depending on the dimension of the supporting affine manifold, pushing us to analyze the cases of dimension two, three and higher, separately. 

In each case, we consider the decomposition of the affine connection compatible with the cosmological and static spherical symmetries, mainly restricting ourselves to the decomposition of the symmetric part of the connection. Interestingly, the transverse-traceless component of the tensor \(S\), denotes \(S^{TT}\), has no proper degrees of freedom. Hence, the geometric meaning of \(S^{TT}\) is ciphered in the choice of the auxiliary metric.

With the decomposed connection, we were able of identify the components of the connection that \emph{control} the conformal part of the geometry (\(W_{\mu} = - V_{\mu}\)), and the one that allows to identify autoparallel curves with geodesic curves (\(V^{\mu} = 0\)). This analysis took us closer to the famous work by Ehlers, Pirani and Schild, in which the implications for the motion of free falling particles and light rays is investigated in the context of General Relativity~\cite{ehlers12_repub_of}.

The issue about the \emph{signature} on affine spaces might be answer, at least in cosmological scenarios, from Eqs.~\eqref{eq:cosmo-relations-2d}, where the metric variables are related to affine variables. Note that such relation can be found only after fixing the gauge redundancy.




Intriguingly, when analyzing the radial geodesics on static spherical scenarios, it is possible to define trapped regions. Therefore, it might be possible to extend the definition of black holes to affine models of gravity.

Additionally, in cosmological models, once the autoparallels have been identified with the geodesics, one can make sense of an affine notion of e-folds, paving the route towards affine inflationary models.

\begin{acknowledgements}
  The authors want to thank Jos\'e Perdiguero for the fruitful discussions during the completion of this article. J.V.-S. acknowledges the Doctoral scholarship of Universidad T\'ecnica Federico Santa Mar\'ia, granted by the Graduate Studies and the incentive program of the Pontificia Universidad Cat\'olica de Valpara\'iso.
\end{acknowledgements}

\bibliographystyle{spphys}
\bibliography{References.bib}

\end{document}